# Lattice Designs in Standard and Simple Implicit Multi-linear Regression

Rebecca D. Wooten

*Department of Mathematics and Statistics, University of South Florida, Tampa 33620*

**Abstract**: Statisticians generally use ordinary least squares to minimize the random error in a subject response with respect to independent explanatory variable. However, Wooten shows illustrates how ordinary least squares can be used to minimize the random error in the system without defining a subject response. Using lattice design Wooten shows that non-response analysis is a superior alternative rotation of the pyramidal relationship between random variables and parameter estimates in multi-linear regression. **Non-Response Analysis** for simple linear co-linearity and **Rotational Analysis** in **Simple Linear Regression** challenge the notion of fixed effects; unity is included as a random measure (variable). The illustrations using lattice designs a **mean operator** that generates the **standard mean** and the **self-weighing mean**, among other point estimates with random weights; and a **join** that illustrates **variance** and **covariance**; and develops the measures of variance referred to as **internal co-variance** and **base variance**. These concepts are used to illustrate how these measures are used to evaluate parameter estimates in standard **simple linear regression** and **simple implicit regression** (**non-response** and **rotational**). The resulting analysis of these lattice designs show standard **simple linear regression** limits the relationship by consider the variance in one direction as relating to the two adjacent **co-variances** (**standard** and **internal**) whereas **non-response analysis** defines the relationship in terms of the **internal co-variances** and the **base variance**.

## 1. Introduction to Simple Linear Regression in Rotation and Non-response Analysis

Lattice Design I & II will be used to address the solutions to the standard univariate model
$$y = \beta$$
and the non-response model
$$1 = \alpha y,$$
where $\alpha = \frac{1}{\beta}$; and the simple linear regression model in standard form
$$y = \beta_0 + \beta_1 x,$$
the rotation of variables to
$$x = \gamma_0 + \gamma_1 y$$
and the non-response model
$$1 = \alpha_1 x + \alpha_2 y.$$

The measures associated with these systems include two variables and unity (1); that is, the general form of these equations require two variables $\{x, y\}$ with constant coefficients and a constant coefficient without movement, $\{1\}$.

Using these three measures, $\{1, x, y\}$, the solutions to these equations can be illustrated using lattice designs; first relating the measures in Lattice Design I and then for a set of data in Lattice Design II.

## 2. Introduction to Lattice Design I & II and the Mean Operator

Consider Lattice Design I for three measures (unity and two variables): $\{1, x, y\}$ where $x$ and $y$ are positive measures. To construct this lattice design, starting with unity in the base level, Figure 1; moving down the lattice by multiply by $x$ if you veer to the left and multiply by $y$ if you veer to the right to create the levels of the lattice.

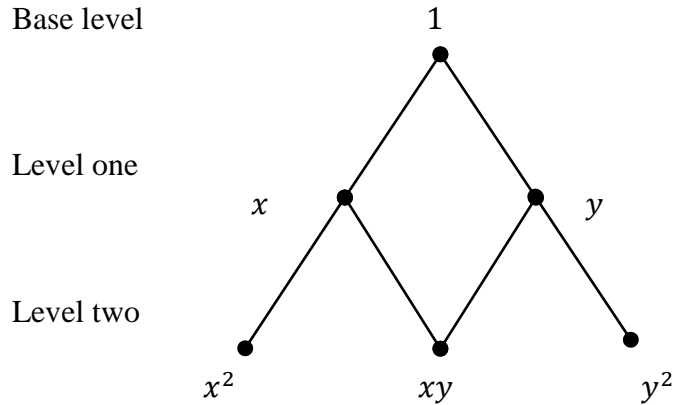

**Figure 1:**   Lattice Design I depicting random variables $x$ and $y$ through level two.

The above lattice represents a single observation. To extend this to a number of observations, let the **product operator** be defined as

$$\prod(a,b) = a \times b$$

and the **additive operator** be defined by

$$\sum(a) = \sum_{i=1}^{n} a_i;$$

then we can defined the **vertices** in the lattice space to be

$$V(a,b) = \sum \prod(a,b).$$

Over a set of data, $\{x_i\}$, this creates Lattice Design II: the base level $V(1,1) = n$ is the sample without variables; the first level or the first order variables: $V(1,x) = \sum x$ and $V(1,y) = \sum y$; and finally the second level with second order terms: $V(x,x) = \sum x^2$, $V(x,y) = \sum xy$ and $V(y,y) = \sum y^2$.

Graphically, this lattice has three levels and six vertices, Figure 2.

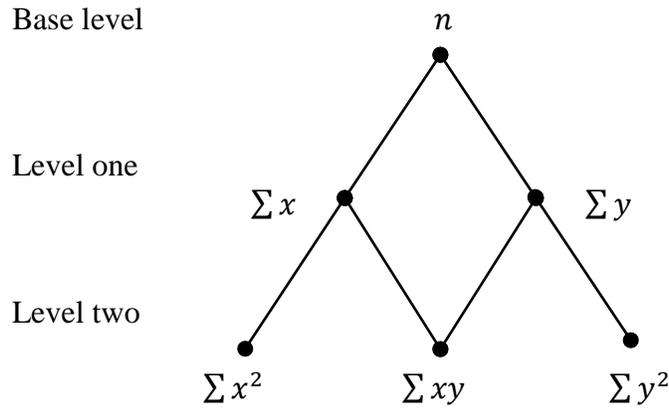

**Figure 2:** Lattice Design II depicting a random set of measures $x_i$ and $y_i$ through level two.

The **means operation** in the direction $d$ from the specified vertex $W = V(a,b)$ is
$$\hat{\mu}_d = M_{Ad} = \frac{\sum abd}{\sum ab} = \frac{\sum Wd}{\sum W}.$$

Steps in computing the means operator:
  Step 1:    Select a starting vertex (weights to be used)
  Step 2:    Designate a direction (variable to be estimated)
  Step 3:    Use the mean operator to estimate the central tendency of the variable

Level-one means are the **standard means**: $\bar{x}$ and $\bar{y}$ with $\bar{x}$ given by
$$\hat{\mu}_x = \frac{\sum 1 \times x}{\sum 1} = \frac{\sum x}{n} = \bar{x}$$
as shown in red, Figure 3.

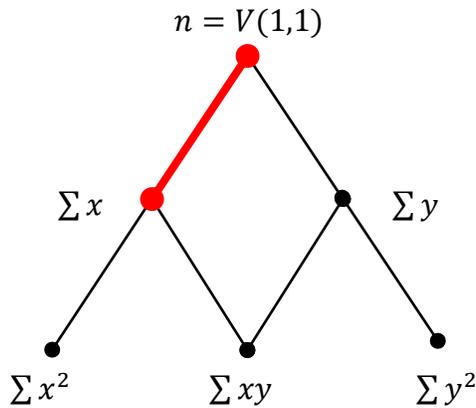

**Figure 3:** Lattice Design II depicting connects made in computing the **standard mean**.

Level-two means in the same direction are the non-response means:
$$\hat{\mu}_x = \frac{\sum x \times x}{\sum x} = \frac{\sum x^2}{\sum x} = \hat{\alpha}$$
the **self-weighting mean** as shown in the illustration below.

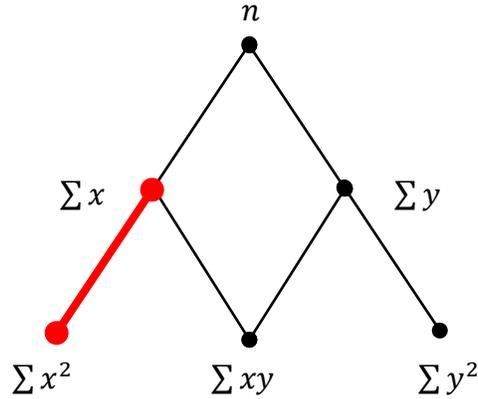

**Figure 4:** Lattice Design II depicting connects made in computing the **self-weighting mean**.

However, using this construct, Figure 5, estimates of $\mu_x$ include randomly weighted means using the measures of $y$ as the weights:
$$\hat{\mu}_x = \frac{\sum x \times y}{\sum y} = \frac{\sum xy}{\sum y}$$

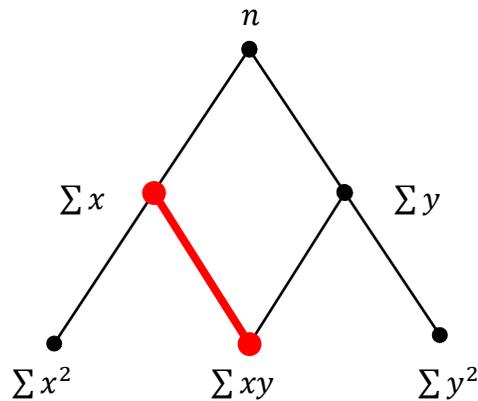

**Figure 5:** Lattice Design II depicting connects made in computing a **randomly weighted mean**.

When $x$ is a normally distributed random variable, any random set of weights will give a comparable measure and for low variance, converge to the standard mean as $\mu_x \to \infty$ irrespective of the sample size $n$.

## 3. Joins and measures of Variance

Let the **product operator** be defined as $\prod(a,b) = a \times b$ and the **additive operator** be defined by $\Sigma(a) = \sum_{i=1}^{n} a_i$; then we can defined the **vertices** in a lattice space to be $V(a,b) = \Sigma \prod(a,b)$ and the **join** between two vertices in two-dimensional space to be

$$W_{abcd} = \prod [V(a,b), V(c,d)],$$

where $a, b, c, d \in \{1, x, y\}$.

Then the **determinates** then take the form of
$$\Delta_{abcd} = W_{abcd} - W_{adcb}.$$

Steps in computing the **join**:
- Step 1: Select two vertices with at least two edges between them
- Step 2: Designate a direction (variable to be estimated)
- Step 3: Use the mean operator to estimate the central tendency of the variable

Lattice Design II also generates **variances**, **covariance** and other measures of deviations by taking the produce of the endpoints and subtracting the middle term squared or subtracting the product of the terms perpendicular to the center.

The first variance, $\sigma_x^2$, is measured in one direction, Figure 6, is centered around $x$ or $\Sigma x$ and is considered a level one measure of variance:

$$\Delta_{11xx} = W[(1,1),(x,x)] - W[(1,x),(1,x)] = n \sum x^2 - \left(\sum x\right)^2 = n^2 \sigma_x^2 = n^2 \sigma_{11xx}^2$$

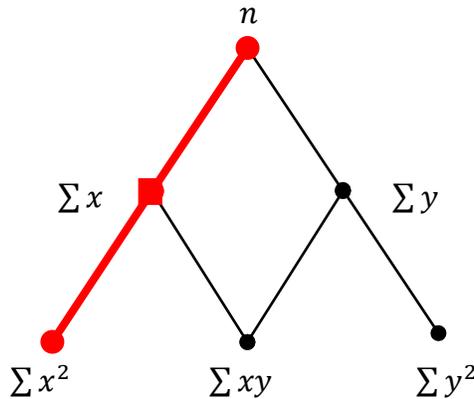

**Figure 6:** Lattice Design II depicting connections made in computing the **variance** of $x$.

The second variance, $\sigma_y^2$, is measured in one direction, Figure 7, is centered around $y$ or $\sum y$ and is considered a level one measure of variance:

$$\Delta_{11yy} = W[(1.1),(y,y)] - W[(1,y),(1,y)] = n\sum y^2 - \left(\sum y\right)^2 = n^2\sigma_y^2 = n^2\sigma_{11yy}^2$$

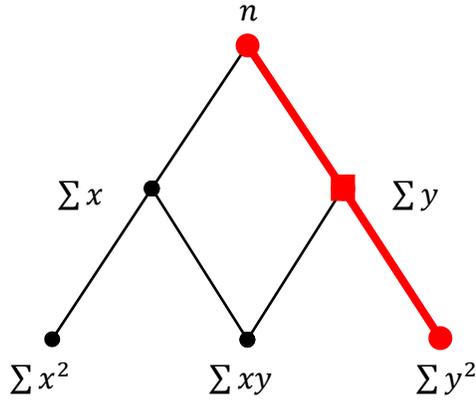

**Figure 7:** Lattice Design II depicting connections made in computing the **variance** of $y$.

The third variance is referred to a **covariance**, $\sigma_{xy}$, is measured between the two directions, Figure 8, is centered between $\sum x$ and $\sum y$ and is considered a level one measure of variance:

$$\Delta_{11xy} = W[(1.1),(x,y)] - W[(1,x),(1,y)] = n\sum xy - \sum x \sum y = n^2\sigma_{xy} = n^2\sigma_{11xy}^2$$

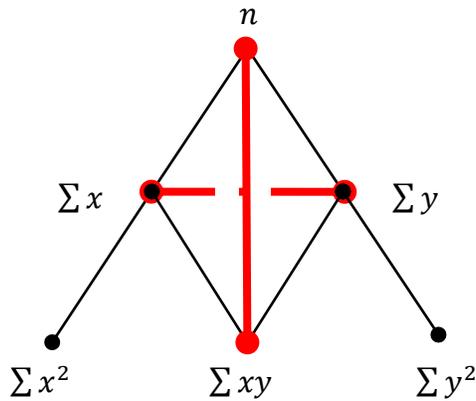

**Figure 8:** Lattice Design II depicting connections made in computing the **co-variance**.

Additional determinates used in standard regression include the following **internal covariance(s)**, Figure 9.

$$\Delta_{1yxx} = W[(1.y),(x,x)] - W[(1,x),(x,y)] = \sum y \sum x^2 - \sum x \sum xy = n^2 \sigma^2_{1yxx}$$

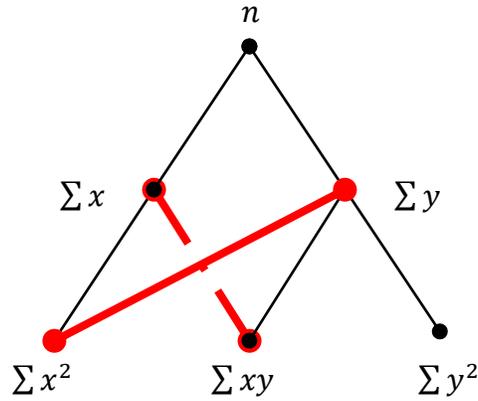

**Fig. 9(a)**

$$\Delta_{1xyy} = W[(1.x),(y,y)] - W[(1,y),(y,x)] = \sum x \sum y^2 - \sum y \sum xy = n^2 \sigma^2_{1xyy}$$

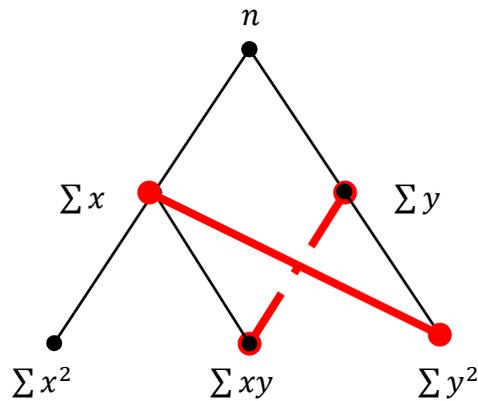

**Fig. 9(b)**

$$\Delta_{1yx1} = W[(1.y),(x,1)] - W[(1,1),(x,y)] = \sum x \sum y - n \sum xy = -\Delta_{11xy}$$

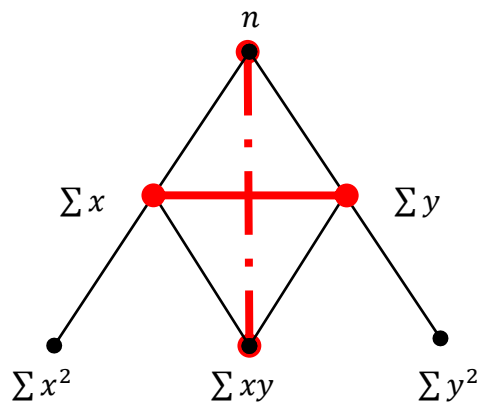

**Fig. 9(c)**
**Figure 9:** Lattice Design II depicting connections made in computing the **internal covariances**.

The determinates are used in standard regression are as follows:

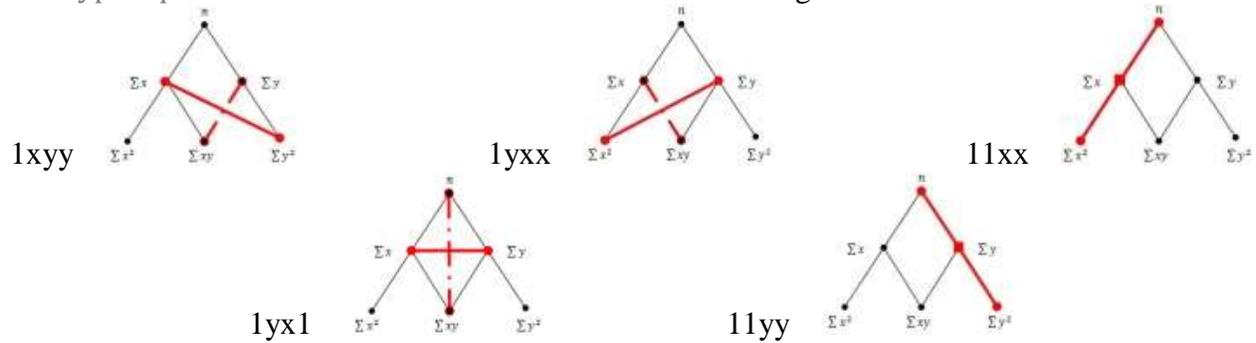

For the standard regression model
$$y = \beta_0 + \beta_1 x,$$
the parameter estimates are given by the

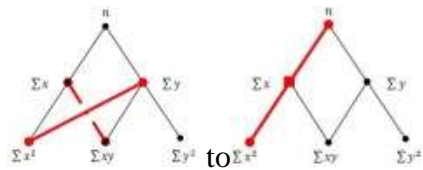

$$\hat{\beta}_0 = \frac{\Delta_{1yxx}}{\Delta_{11xx}}$$

and

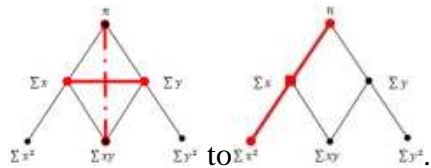

$$\hat{\beta}_1 = \frac{\Delta_{1yx1}}{\Delta_{11xx}}$$

For the standard regression model with rotation of variables
$$x = \gamma_0 + \gamma_1 y$$
the parameter estimates are given by the

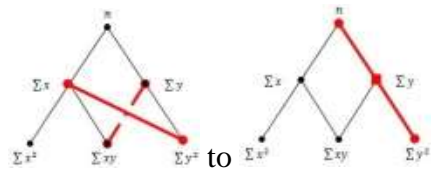

$$\hat{\gamma}_0 = \frac{\Delta_{1xyy}}{\Delta_{11yy}}$$

and

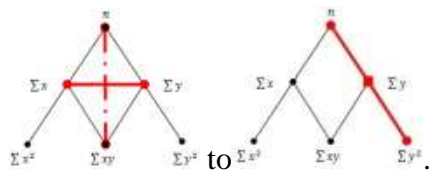

$$\hat{\gamma}_1 = \frac{\Delta_{1xy1}}{\Delta_{11yy}}$$

In the first set of parameter estimates for $y = \beta_0 + \beta_1 x$, the solutions (parameter estimates) depend on the sample size, $n$, and $\sum y^2$. In the second set of parameter estimates for $x = \gamma_0 + \gamma_1 y$, depends on the sample size, $n$, and in this rotation, $\sum x^2$. That is, standard regression limits the relationship by consider the variance in one direction as related to the two adjacent covariances.

The remaining determinate is the **base variance**, $\omega_{xy}^2$, or **second order covariance**; a measure of the variance in level two, Figure 10, is given by

$$\Delta_{xxyy} = W[(x.x),(y,y)] - W[(x,y),(y,x)] = \sum x^2 \sum y^2 - \left(\sum xy\right)^2.$$

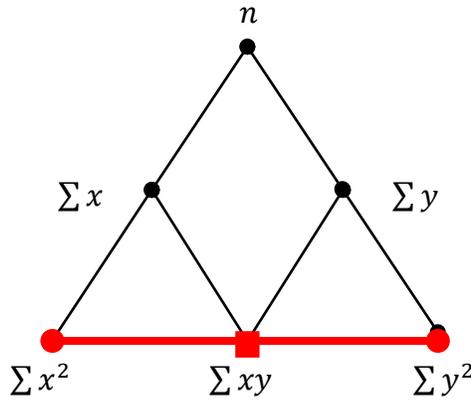

**Figure 10:** Lattice Design II depicting connects made in computing the **base variance**.

This determinate is used in **non-response analysis** to estimate the coefficients that drive the relationship between $x$ and $y$ irrespective of the sample size, $n$:

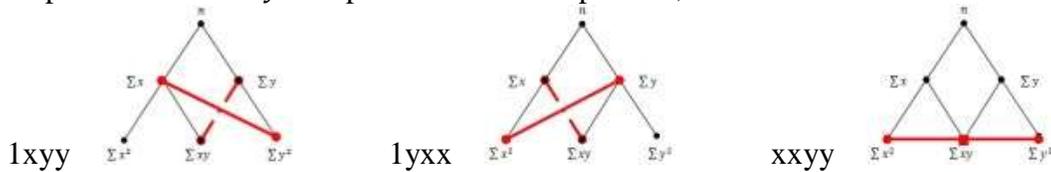

For the **non-response** model

$$1 = \alpha_1 x + \alpha_2 y$$

the parameter estimates are given by the

$$\hat{\alpha}_1 = \frac{\Delta_{1xyy}}{\Delta_{xxyy}}$$

$\sum y^2$ to $\sum x^2$

and

$$\hat{\alpha}_2 = \frac{\Delta_{1yxx}}{\Delta_{xxyy}}$$

$\sum y^2$ to $\sum x^2$.

These solutions are not defined in terms of the sample size $n$, but rather depends on $\sum y^2$ and $\sum x^2$. That is, using ordinary least squares and regression to the means, non-response analysis defines the relationship by considering the base variance between the two directions as related to the two adjacent internal co-variances.

**4.**  **Rotational analysis in three dimensional planes**

Rotational analysis uses ordinary least squares and standard regression methods to test the relationship among a set of terms by rotating the terms into the response variable one at a time.

Consider the three-dimensional plane defined by
$$d = \alpha_1 a + \alpha_2 b + \alpha_3 c,$$
where $a, b, c, d$ are unique elements from the observed space and unity; $a, b, c, d \in \{1, x, y, z\}$.

$$\begin{bmatrix} d_1 \\ \vdots \\ d_n \end{bmatrix} = \begin{bmatrix} a_1 & b_1 & c_1 \\ \vdots & \vdots & \vdots \\ a_n & b_n & c_n \end{bmatrix} \begin{bmatrix} \alpha_1 \\ \alpha_2 \\ \alpha_3 \end{bmatrix}$$

The first three rotations are standard regression and if $d = 1$, then this is non-response analysis; however, the solutions to the normal equations are all related to the determinants as defined by Cramer's rule.

Give the following normal equations:

$$\begin{bmatrix} \sum a^2 & \sum ba & \sum ca \\ \sum ab & \sum b^2 & \sum cb \\ \sum ac & \sum bc & \sum c^2 \end{bmatrix} \begin{bmatrix} \alpha_1 \\ \alpha_2 \\ \alpha_3 \end{bmatrix} = \begin{bmatrix} \sum ad \\ \sum bd \\ \sum cd \end{bmatrix},$$

then define the main matrix and three augmented matrixes to be

$$W = \begin{bmatrix} \sum a^2 & \sum ba & \sum ca \\ \sum ab & \sum b^2 & \sum cb \\ \sum ac & \sum bc & \sum c^2 \end{bmatrix},$$

$$W_1 = \begin{bmatrix} \sum ad & \sum ba & \sum ca \\ \sum ad & \sum b^2 & \sum cb \\ \sum ad & \sum bc & \sum c^2 \end{bmatrix}, W_2 = \begin{bmatrix} \sum a^2 & \sum ad & \sum ca \\ \sum ab & \sum bd & \sum cb \\ \sum ac & \sum cd & \sum c^2 \end{bmatrix} \text{ and } W_3 = \begin{bmatrix} \sum a^2 & \sum ba & \sum ad \\ \sum ab & \sum b^2 & \sum bd \\ \sum ac & \sum bc & \sum cd \end{bmatrix}.$$

Then the parameter estimates can be determined using determinates: $\alpha_i = \frac{|W_i|}{|W|}$.

Consider an extension of Lattice Design II to Lattice Design III, illustrated in Figure 11, to include $\{1, x, y, z\}$. This construct is equivalent to a pyramid that can be rotated to focus in on variable at a time where unity (1) is held fixed at the top in standard regression and the focus is on $y$, Figure 11.

This figure contains all the variances, co-variances (level one and level two) as well as ternary-variances for second-order interactions.

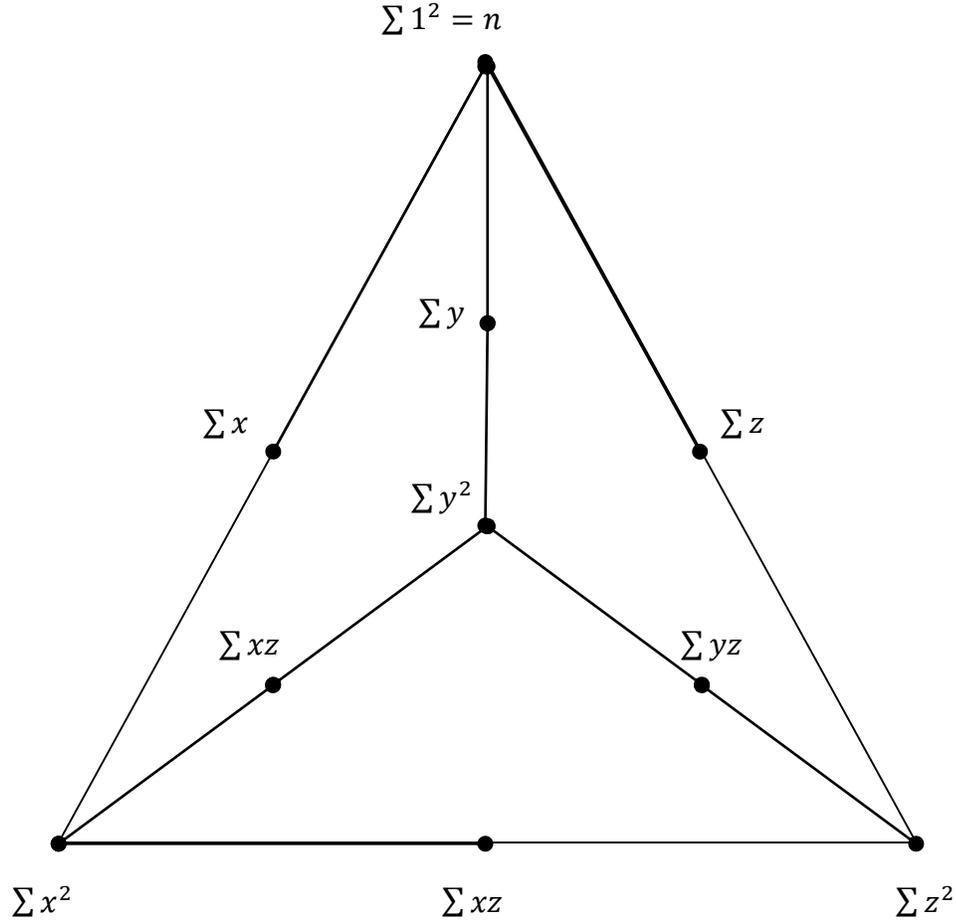

**Figure 11:** Lattice Design III depicting random set of measures $x_i$, $y_i$ and $z_i$ through level two.

There are four rotations which can be generalize using arbitrary directions $\{a, b, c, d\}$ over the fixed view of $\{1, x, y, z\}$. Recall the product operator be defined as $\prod(a, b) = a \times b$ and the additive operator be defined by $\Sigma(a) = \sum_{i=1}^{n} a_i$; then we can defined the vertices in a lattice space to be $V(a, b) = \Sigma \prod(a, b) = ab$; and define the **join** between three vertices in three-dimensional space to be

$$W[(a,b),(c,d),(e,f)] = \prod [V(a,b), V(c,d), V(e,f)],$$

where $a, b, c, d, e, f \in \{1, x, y, z\}$.

Then the determinates used to evaluate the parameter estimates take the form of

$$\begin{aligned}\Delta_{abcdef} = &W[(a,b),(c,d),(e,f)] - W[(a,b),(c,f),(e,d)] \\ &-W[(a,d),(c,b),(e,f)] + W[(a,d),(c,b),(e,f)] \\ &+W[(a,b),(c,f),(e,d)] - W[(a,b),(c,d),(e,f)].\end{aligned}$$

$$\sum a^2$$

$$\sum ad$$

$$\sum ab \qquad \sum ac$$

$$\sum d^2$$

$$\sum bd \qquad \sum cd$$

$$\sum b^2 \qquad \sum bc \qquad \sum c^2$$

**Figure 12:** Lattice Design III depicting random set of measures $\{a, b, c, d\}$ over the fixed view of $\{1, x, y, z\}$ through level two.

Determinates needed to estimate the parameters in the rotational model $d = \alpha_1 a + \alpha_2 b + \alpha_3 c$ take the following form:

| Focus: | $y$ | $x$ | $z$ | $1$ |
|---|---|---|---|---|
| Denominator ($\Delta$) $\Rightarrow$ | $\Delta_{11xxzz}$ | $\Delta_{11yyzz}$ | $\Delta_{11xxyy}$ | $\Delta_{xxyyzz}$ |
| Numerator ($\Delta\hat{\alpha}$) $\Downarrow$ | | | | |
| $\theta_1 = \Delta\hat{\alpha}_1$ | $\Delta_{1yxxzz}$ | $\Delta_{1xyyzz}$ | $\Delta_{1zxxyy}$ | $\Delta_{x1yyzz}$ |
| $\theta_2 = \Delta\hat{\alpha}_2$ | $\Delta_{11xyzz}$ | $\Delta_{11yxzz}$ | $\Delta_{11xzyy}$ | $\Delta_{xxy1zz}$ |
| $\theta_3 = \Delta\hat{\alpha}_3$ | $\Delta_{11xxzy}$ | $\Delta_{11yyzx}$ | $\Delta_{11xxyz}$ | $\Delta_{xxyyz1}$ |

The determinant in the denominator and numerator take one of two distinct forms, Form I and Form II, respectively.

Form 1: $\quad \Delta_{aabbcc} = W[(a,a),(b,b),(c,c)] - W[(a,a),(b,c),(c,b)]$
$\quad\quad\quad\quad\quad - W[(b,a),(b,a),(c,c)] + W[(b,a),(b,c),(c,a)]$
$\quad\quad\quad\quad\quad + W[(c,a),(b,a),(c,b)] - W[(c,a),(b,b),(c,b)].$

Representative joins for the positive terms in Form I:
$W[(a,a),(b,b),(c,c)] \quad\quad\quad W[(b,a),(b,c),(c,a)] \quad\quad\quad W[(c,a),(b,a),(c,b)]$

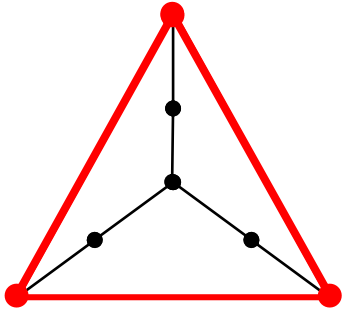
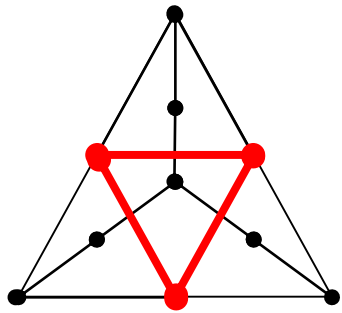
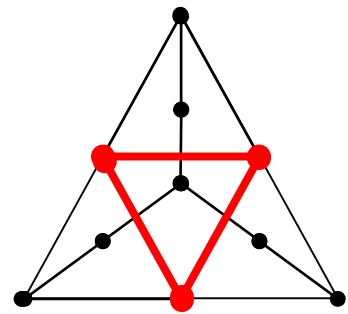

**Fig.13(a)**            **Fig.13(b)**            **Fig.13(c)**

Representative joins for the negative terms in Form I:
$W[(a,a),(b,c),(c,b)] \quad\quad\quad W[(b,a),(b,a),(c,c)] \quad\quad\quad W[(c,a),(b,b),(c,b)]$

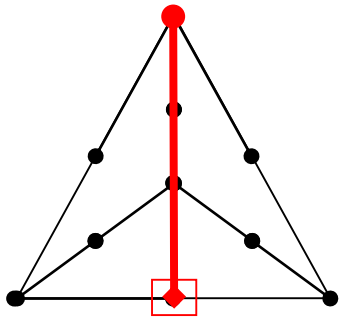
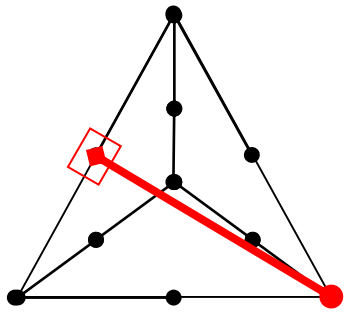
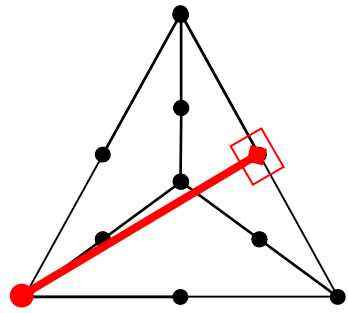

**Fig.13(d)**            **Fig.13(e)**            **Fig.13(f)**

**Figure 13:**     Lattice Design III depicting connects made in computing the **determinate** of the form $\Delta_{aabbcc}$.

FORM II:  $\Delta_{adbbcc} = W[(a,d),(b,b),(c,c)] - W[(a,d),(b,c),(c,b)]$
$\phantom{\text{FORM II: }\Delta_{adbbcc} = }-W[(b,a),(b,d),(c,c)] + W[(b,a),(b,c),(c,d)]$
$\phantom{\text{FORM II: }\Delta_{adbbcc} = }+W[(c,a),(b,d),(c,b)] - W[(c,a),(b,b),(c,d)].$

Representative joins for the positive terms in Form II:
$W[(a,d),(b,b),(c,c)]$    $W[(b,a),(b,c),(c,d)]$    $W[(c,a),(b,d),(c,b)]$

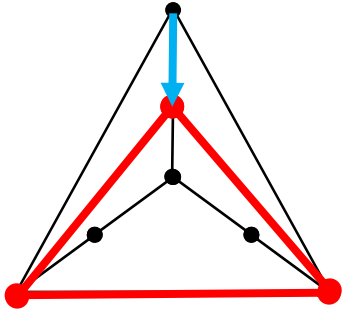 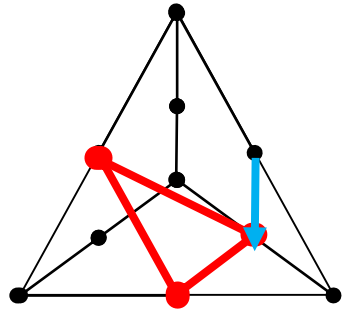 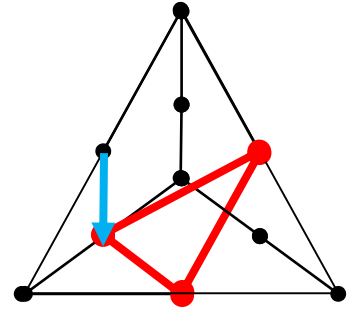

**Fig.14(a)**    **Fig.14(b)**    **Fig.14(c)**

Representative joins for the negative terms in Form II:
$W[(a,d),(b,c),(c,b)]$    $W[(b,a),(b,d),(c,c)]$    $W[(c,a),(b,b),(c,d)]$

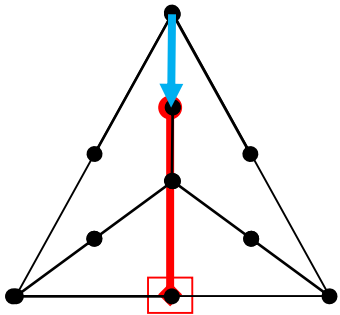 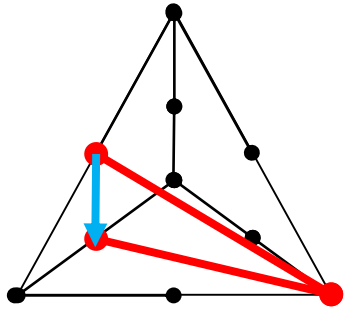 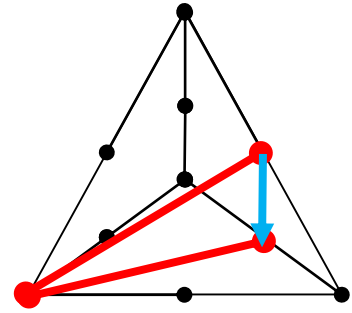

**Fig.14(d)**    **Fig.14(e)**    **Fig.14(f)**

**Figure 14:**  Lattice Design III depicting connects made in computing the **determinate** of the form $\Delta_{adbbcc}$.

Form II is Form I suppressed in the $d$ direction. Therefore, there is a close relationship between standard multi-linear regression and non-response analysis in construction, differing only by the focus $x, y, z$ or unity.

## 5.  Conclusion

There are many common weighted means; using non-response analysis we obtain the self-weighting mean and using the mean operator we can also define the randomly weighted mean. Using Cramer's Rule, we can solve the system of equation resulting for rotating each variable term into the subject response position including unity as a random variable in order to hone in on the true linear relationship and focus in on the error in the system in each direction and to the plane.

Non-response analysis focuses in on the error in the system and not in any one direction; treating unity, the constant in the system, as a random effect.

This type of analysis can be extended into such statistical models as
$$1 = \alpha_1 x + \alpha_2 y + \alpha_3 xy$$
and other implicit functions including higher order terms. The usefulness of which is subject to the tractability of the variable measures.

## References

Cramer, G. (1750), *Intr. à l'analyse de lignes courbes algébriques*. Geneva, 657-659, 1750.

Jamison, B., Orey, S., and Pruitt, W., *Convergence of Weighted Averages of Independent Random Variables,* Zeitschrift für Wahrscheinlichkeitstheorie und Verwandte Gebiete, 1965, Volume 4, Issue 1, pp 40-44

Muir, T. *The Theory of Determinants in the Historical Order of Development*, Vol. 1. New York: Dover, pp. 11-14, 1960.

Nasrollah Etemadi, *Convergence of Weighted Averages of Random Variables Revisited*, Proceedings of the American Mathematical Society, Volume 134, Number 9, September 2006, Pages 2739–2744, S 0002-9939(06)08296-7 Article electronically published on April 10, 2006

Norman L. Johnson & Samuel Kotzb, *Randomly Weighted Averages: Some Aspects and Extensions*, The American Statistician, Volume 44, Issue 3, 1990, pages 245-249, DOI:10.1080/00031305.1990.10475730

Weisstein, Eric W. "*Cramer's Rule*." From MathWorld--A Wolfram Web Resource, 1999-2015, http://mathworld.wolfram.com/CramersRule.html

Wooten, R. D., Baah, K., DAndrea, J. (2015) *Implicit Regression: Detecting Constants and Inverse Relationships with Bivariate Random Error*, Cornell University Library, arXiv:1512.05307

Wooten, R. D., (2016) *An Introduction to Implicit Regression: Extending Standard Regression to Rotational Analysis and Non-Response Analysis*, Cornell University Library, arXiv:1602.00158

Wooten, R. D., *Statistical Analysis of the Relationship between Wind Speed, Pressure and Temperature*, Journal of Applied Sciences, 2011, ISSN 1812-5654 / DOI: 10.3923/jas.2011., Asian Network for Science Information